# Improved Complexity Bound of Vertex Cover for Low degree Graph

Weiya Yue[1], John Franco[1], Weiwei Cao[1,2]

[1] Computer Science Department, University of Cincinnati, Ohio, US
[2] State Key Laboratory of Information Security, Graduate University of Chinese Academy of Sciences, Beijing, China
`weiyayue@hotmail.com,franco@gauss.ececs.uc.edu,wwcao@is.ac.cn`

**Abstract.** In this paper, we use a new method to decrease the parameterized complexity bound for finding the minimum vertex cover of connected max-degree-3 undirected graphs. The key operation of this method is reduction of the size of a particular subset of edges which we introduce in this paper and is called as "real-cycle" subset. Using "real-cycle" reductions alone we compute a complexity bound $O(1.15855^k)$ where $k$ is size of the optimal vertex cover. Combined with other techniques, the complexity bound can be further improved to be $O(1.1504^k)$. This is currently the best complexity bound.

## 1 Introduction

Given an undirected connected graph $G = (V, E)$, a vertex cover of $G$ is a subset $V'$ of $V$ such that for every edge $(x, y)$ in E, at least one of $x$ or $y$ is in $V'$. The problem of finding a minimum cardinality vertex cover is a classical optimization problem that has numerous applications and has been studied for decades. For example, in the field of bioinformatics, it can be used in the construction of phylogenetic trees, in phenotype identification, and in analysis of microarray data. Its decision version, known as the Vertex Cover Problem (VC), is: given graph $G$ and an integer $k$, decide whether $G$ has a vertex cover of $k$ vertices. The Vertex Cover Problem is one of Karp's 21 original NP-complete problems [7].

Fixed Parameter Tractability (FPT) is a framework that attempts to cope with and understand the complexity limitations of NP-complete problems. In [1] it is proved that VC is FPT. The fixed parameter complexity of VC takes the form $f(k)n^{O(1)}$ in which $k$ is the cardinality of vertex cover and $f$ is a function that depends only on $k$. Then, for fixed $k$, VC can be solved in polynomial time. In [2] the fixed parameter complexity of VC is shown to be $O(kn + 2^k k^{2k+2})$. This is improved to be $O(kn + 2^k k^2)$ in [3]. By further analysis, Niedermeier and Rossmanith developed an algorithm of complexity $O(kn+1.2918^k k^2)$ in [4]. This is improved to be $O(kn + 1.271^k k^2)$ in [5]. For most practical problems $k$ is quite large so, although the above theoretical results are impressive, they are not useful for many real problems.

It has been pointed out that some real VC problems involve graphs of low degree [5,4]. Fixed Parameter Tractability of VC on graphs of maximum degree

3 has been studied to determine whether low degree can be exploited. The problem of finding a minimum cardinality vertex on such graphs is abbreviated vc-3. In [6] a $O(k^2 1.194^k + n)$ bound on the fixed parameter complexity of vc-3 is shown. In [8] this is improved to $O(1.1864^k + n)$. The best published result of $O(1.1849^k)$ appears in [11]. In that paper it is said that a recent unpublished technical report [12] by Bourgeois et al. claims that the Maximum Independent Set problem on undirected graphs of degree 3 can be solved in $O(1.0854^n)$ time. Since the Maximum Independent Set problem is the dual of VC, this implies a complexity of $O(1.1781^k)$ for vc-3. Our contribution to reduce the fixed parameter complexity of vc-3 to $O(1.1710^k)$. This is better than the best published or reported results and uses the notion of "real-cycle."

The notion of "real-cycle" is the key to our analysis. All current algorithms for VC are based on the branch-and-bound search paradigm. We investigate an algorithm that takes a different approach by iteratively decreasing the cardinality of a real-cycle subset.

In Section 2, we define real-cycles, present other needed terminology, and prove some facts that will be needed later. In Section 3, we describe the basic mechanism our algorithm uses to solve VC. The difference between this mechanism and current approaches to solving VC is described as well. In Section 4, our algorithm is presented as pseudo code and described. In Section 5, the algorithm's fixed parameter complexity is analyzed. Section 6 summarizes the results and suggests the next steps in this work.

## 2 Terminology and Foundation

Conventional graph theory terminology is applied here. Graph $G = (V, E)$ is a undirected graph with vertex set $V$ and edge set $E$. Letters $n$ and $m$ represent the cardinality of $V$ and $E$, respectively. Letters $u, v$ and $w$ represent *vertices*. An edge is a set of two vertices, for example $\{u, v\}$. Denote by $deg(u)$ the degree of vertex $u$. Denote by $N(u)$, called the neighborhood of $u$, the vertices of $G$ that are directly connected to $u$ by an edge in $E$. A simple cycle in $G$ is a cycle with no loops. Let $U \subset V$ be a non-empty subset of vertices in $G$. Define $G'(U, G) = (V \setminus U, \{e \in E : \forall v \in U, \ v \notin e\})$ to be a subgraph of $G$ that is reduced by eliminating all vertices in $U$ and edges in $E$ that are incident to at least one of them. Denote the cardinality of set $S$ by $|S|$. Denote by $VC(G)$ the minimum vertex cover of a given graph $G$.

Next, we give definitions of the key notions real-cycle and extra degree of a graph. Then we prove some lemmas and properties to prepare for the analysis of our algorithm to solve *vc*-3 in section 4.

**Definition 1.** *Let $G$ be an undirected graph. A* real-cycle subset *of $G$ is a list $[c_1, c_2, ..., c_l]$ of simple cycles in $G$ such that, $\forall_{1 < i \leq l} \ \exists e \in c_i \ \forall_{1 \leq j < i} \ e \notin c_j$*

In words, every simple cycle $c$ of a real-cycle subset of $G$ has at least one edge that is not a member of any simple cycle that is listed prior to $c$ in the real-cycle subset of $G$.

It is evident that if some cycles are deleted from a real-cycle subset, what remains is still a real-cycle subset. If one real-cycle subset has maximum cardinality among all real-cycle subsets, then it is called a *maximum real-cycle subset* and is denoted as max-$RC(G)$. Note that there may be more than one maximum real-cycle subset. Every $c \in RC(G)$ is said to be a *real-cycle* with respect to $RC(G)$. The cardinality of $RC(G)$ is called the *real-cycle number*. Denote $\tau(G) = |\text{max-}RC(G)|$.

**Definition 2.** *Let $G = (V, E)$ be an undirected graph. The* extra-degree *of a vertex $v \in V$, denoted $ex(v)$, is the maximum of 0 and the degree of $v$ minus 2. The* extra degree *of graph $G$, denoted $ex(G)$, is $ex(G) = \sum_{v \in V} ex(v)$.*

**Definition 3.** *Let $G$ be an undirected graph. A* line *$\mathcal{L} \in G$ is a path through $G$ starting from a vertex whose degree is 1, spanning vertices whose degree is 2, and ending at a vertex whose degree is not 2.*

**Lemma 1.** *Let $G = (V, E)$ be a connected undirected graph. Let $v$ be a degree 1 vertex in $G$. Let $\mathcal{L}$ be a line starting from $v$ and ending at $w$. If $deg(w) = 1$, delete from $G$ all vertices and edges incident to vertices along $\mathcal{L}$; otherwise, if $deg(w) \geq 3$, delete from $G$ all vertices and edges incident to all vertices along $\mathcal{L}$ except $w$. Let $G'$ be the graph that remains. Then $\tau(G') = \tau(G)$. Moreover, if $G'$ is empty then $ex(G') = ex(G)$; otherwise $ex(G') = ex(G) - 1$.*

*Proof.* By definition, a line cannot exist in any cycle. Hence, deletion of a line does not affect a given real-cycle subset and $\tau(G') = \tau(G)$.

It is evident that $G'$ is empty if and only if $deg(w) = 1$ and $G$ is a one-line graph (in fact, it is $\mathcal{L}$). By definition of $ex$, only vertices whose degrees are $> 2$ can change the $ex$ value of a graph if their degrees are changed. Since all vertices on $\mathcal{L}$ have degree $\leq 2$, $ex(G') = ex(G)$. If $G'$ is not empty, since $deg(w) \geq 3$, and $ex(G)$ is decreased by 1 since $deg(w)$ is decreased by 1. Hence, if $G'$ is not empty, $ex(G') = ex(G) - 1$. □

**Lemma 2.** *Let $G = (V, E)$ be a connected undirected graph. Suppose $\{u, v\} \notin E$ and let $G' = G(V, E \cup \{\{u, v\}\})$. Then $\tau(G') = \tau(G) + 1$.*

*Proof.* Since $G$ is connected, there must be more than one path between $u$ and $v$. Arbitrarily choose one and call it $p$. Add edge $e = \{u, v\}$ to $G$ to get $G'$. Path $p$ plus $e$ forms a simple cycle in $G'$ and that simple cycle could not be a member of any max-$RC(G)$. Moreover, $e$ is an edge that does not exist in any simple cycle of max-$RC(G)$. Hence, $p$ plus $e$ may be added to max-$RC(G)$ to get a real-cycle subset of $G'$ and $\tau(G') \geq \tau(G) + 1$.

Next, choose any max-$RC(G')$ and call it $R$. Suppose there are $x$ real-cycles in $R$ containing $\{u, v\}$. Observe that $x \geq 1$ because, if $x = 0$, then the simple cycle $p$ plus $e$ can be subsequently added to $R$ contradicting the hypothesis that $R$ is a real-cycle subset of maximum cardinality.

Name the $x$ real-cycles following the order they appear in $R$ as $c_{\pi_1}, c_{\pi_2}, ..., c_{\pi_x}$ and without loss of generality assume $c_{\pi_1}$ is the real-cycle which uses $e = \{u, v\}$ as the reason (as the non-shared edge) to be a real-cycle. By definition, each cycle in $c_{\pi_2}, ..., c_{\pi_x}$ has at least one edge which is not shared with lower indexed real-cycles. Name one of the edges that is not shared with lower-indexed real-cycles corresponding to $c_{\pi_2}, ..., c_{\pi_x}$ as $e_2, ..., e_x$, respectively. Let $e_i = \{a_i, b_i\}$, $2 \leq i \leq x$.

We need to show that every $c_{\pi_i}$, $2 \leq i \leq x$, in $R$ can be replaced by a real-cycle that does not contain $\{u, v\}$ but still uses $e_i$ as the reason to be a real-cycle in $R$. Then, if $c_{\pi_1}$ is deleted from $R$, $R$ becomes a maximum cardinality real-cycle subset for $G$. Thus, $\tau(G') - 1 \leq \tau(G)$ and with the above result we get $\tau(G') = \tau(G) + 1$.

Assume $c_1 = \{u \overset{p}{-} v \overset{e}{-} u\}$ and $c_i = \{u \overset{p_{i1}}{-} a_i \overset{e_i}{-} b_i \overset{p_{i2}}{-} v \overset{e}{-} u\}$, and $e_i \notin c_1$, $2 \leq i \leq x$. Then between $a_i, b_i$ there is another path induced from $a_i \overset{p_{i1}}{-} u \overset{p}{-} v \overset{p_{i2}}{-} b_i$. Thus $a_i \overset{p_{i1}}{-} u \overset{p}{-} v \overset{p_{i2}}{-} b_i \overset{e_i}{-} a_i$ is a new real cycle that does not contain $e = (u, v)$ but still uses $e_i$ as the reason to be a real-cycle in max-$RC(G')$ and therefore it can replace $c_i$. □

Equivalently, if an edge is deleted from connected graph $G$, and the resulting graph $G'$ is connected, then by Lemma 2, $\tau(G') = \tau(G) - 1$.

**Theorem 1.** *Let $G = (V, E)$ be a connected undirected graph. If the minimum vertex degree of $G$ is at least 2, then $\tau(G) = \frac{ex(G)}{2} + 1$.*

*Proof.* Let $|V| = n$. We prove the theorem by induction on $n$. Since the minimum vertex degree of $G$ is $\geq 2$, then it is necessary that $n \geq 3$. When $n = 3$, there is only one real-cycle, and $|\text{max-}RC(G)| = 1$. Thus $\tau(G) = \frac{ex(G)}{2} + 1$ holds when $n = 3$.

Let $q$ be an integer greater than 3 and suppose the hypothesis holds for $3 \leq n < q$. Consider $G$ with $n = q$ vertices. Choose one vertex $v$ arbitrarily and remove $v$ and its incident edges, one at a time, to get $G'$. Let $w$ be a neighbor of $v$ in $G$. Let $G_w$ be the graph before the edge $\{v, w\}$ is deleted. Let $G'_w$ be the graph after $\{v, w\}$ is deleted. Consider three cases (for convenience, in these cases we temporarily do not consider the change of $ex$ value caused by $v$ and count it later in the proof):

1. $G'_w$ is connected and $deg(w) = 1$ (then in $G_w$ $deg(w) = 2$). In this case a line $\mathcal{L}$ appears in $G'_w$ starting at $w$. Delete the vertices and their incident edges along $\mathcal{L}$ from $w$ to a vertex $v'$ such that a) $deg(v') \geq 2$ or b) $v' = v$. Call the remaining graph $G'''_w$ For subcase a), by Lemma 2 and Lemma 1, $\tau(G'''_w) = \tau(G_w) - 1$ and $ex(G'''_w) = ex(G_w) - 1$. For subcase b) $\mathcal{L}$ passes through a cycle, and since we do not consider the change of $ex$ value caused by $v$, we have $ex(G'''_w) = ex(G_w)$ and by Lemma 2 and Lemma 1 $\tau(G'_w) = \tau(G_w) - 1$.
2. $G'_w$ is connected and $deg(w) > 1$ (then in $G_w$ $deg(w) > 2$). In this case, by Lemma 2, $\tau(G'_w) = \tau(G_w) - 1$ and $ex(G'_w) = ex(G_w) - 1$ since $deg(w)$ is decreased by 1. The change is the same as was discussed in subcase a) above. Assume the change in case 1a) and 2) appear $x$ times and case 1b) appear $y$ times after deleting all $v$'s incident vertices and $v$.

3. $G'_w$ is disconnected and it consist of two isolated subgraphs $G_1$ and $G_2$. Assume $v \in G_1$ and $w \in G_2$. If $deg(w) = 1$, then remove from $G_2$ all vertices and their incident edges from the line starting at $w$. Let $G'_2$ be remaining graph. Since $G'_2$ is not a one-line graph (if it is, it contradicts the hypothesis that the minimum vertex degree of $G$ is at least 2), by Lemma 1 $ex(G'_w) = ex(G_w) - 1$. Moreover, since $G_1$ and $G_2$ are disconnected, the edge $e$ cannot exist in any simple cycle? Hence $\tau(G'_w) = \tau(G_w)$.

Assume the change in case 3) appear $z$ times after deleting all $v$'s incident vertices and $v$.

Note that the last deleted incident edge of $v$ must cause one subgraph $G_1 = \{v\}$ which can be discarded as isolated vertex directly. So in $G'$ there are $z$ connected components with minimum degree $\geq 2$. Since for $n < q$ the equations hold for each component, then by summation $\tau(G') = \frac{ex(G')}{2} + z$.

In case 1a), 2) and 3) only one incident edge of $v$ is deleted and in case 1b) two edges of $v$ are deleted altogether, hence $deg(v) = 2x + y + z$. From the $\tau$ change analysis case by case, we see that $\tau(G) = \tau(G') + x + y = \frac{ex(G')}{2} + x + y + z$ and $ex(G) = ex(G') + (deg(v) - 2 + y + z) = ex(G') + 2x + 2y + 2z - 2$, then $\frac{ex(G)}{2} + 1 = \frac{ex(G')}{2} + x + y + z = \tau(G)$. □

Note that $ex(G)$ must be an even number. Since the minimum vertex degree is $\geq 2$, then $ex(G) = \sum_v (deg(v) - 2) = 2m - 2n$ where $m = |E|, n = |V|$.

**Theorem 2 (YC Theorem).** *Let $G = (V, E)$ be a connected undirected graph. A max-RC(G) contains at most* $\text{floor}(\frac{ex(G)}{2}) + 1$ *real-cycles, i.e.*

$$\tau(G) \leq \text{floor}(\frac{ex(G)}{2}) + 1.$$

*Here* $\text{floor}(q)$ *means the maximum integer less or equal to $q$.*

*Proof.* Let $G'$ be obtained by deleting all lines in $G$ as in Lemma 1. Then $G'$ is empty or is connected and has minimum degree $\geq 2$. By Lemma 1, $ex(G') \leq ex(G)$ and $\tau(G') = \tau(G)$. Then $\tau(G) = \frac{ex(G')}{2} \leq \frac{ex(G)}{2}$. Since $ex(G)$ may be an odd number, hence $\tau(G) \leq floor(\frac{ex(G)}{2}) + 1$. □

**Corollary 1 (YC Corollary).** *Let $G = (V, E)$ be a connected undirected graph. If for all $v \in V$, $deg(v) \geq 3$, then $\tau(G) = floor(\frac{n}{2}) + 1$.*

*Proof.* Because all vertices of $G$ have degree at least 3, $ex(G) = \sum_v (deg(v) - 2) \leq n(3 - 2) = n$. Then, by Theorem 2, $\tau(G) \leq floor(\frac{n}{2}) + 1$. □

**Proposition 1.** *Let $G$ be a connected undirected graph. If $\tau(G) = 0$, $G$ is a tree.*

*Proof.* If max-$RC(G)$ is empty, then there is no cycle. □

```
        VC(V, E, k) =
1.        If k ≤ 0, return true.
2.        If V = ∅, return false.
3.        v, V', E' := select(V, E).
4.        Let E''_1 := {e ∈ E' : e is not incident to v}.
5.        If (V' \ {v}, E''_1) is a tree, set n_1 := VC_{poly}(V' \ {v}, E''_1, k − 1),
6.        Otherwise, set n_1 := VC(V' \ {v}, E''_1, k − 1).
7.        Let V'' := {w ∈ V' : {w, v} ∈ E'}.
8.        Let E''_2 := {e ∈ E' : e is not incident to a vertex in V''}.
9.        If (V' \ (V'' ∪ {v}), E''_2) is a tree,
10.          set n_2 := VC_{poly}(V' \ (V'' ∪ {v}), E''_2, k − |V'|).
11.       Otherwise, set n_2 := VC(V' \ (V'' ∪ {v}), E''_2, k − |V'|).
12.       Return n_1 ∨ n_2.
```

**Fig. 1.** Divide and conquer algorithm for vertex cover

## 3   Mechanism and Comparison with Prior Work

The vertex cover algorithm we analyze, called $VC$ here, is just a the standard, simple, well-known, divide-and-conquer algorithm with the modification that if a subgraph is a tree, then its minimum vertex cover can be solved directly with a polynomial time algorithm. Let $VC_{poly}$ denote such a polynomial time algorithm that takes the same inputs and produces the same output as $VC$. Algorithm $VC$ is shown in Figure 1. This simplified version answers the question whether there exists a cover of size $k$ or less but does not return such a cover.

This algorithm will be analyzed assuming a novel strategy for selecting vertex $v$ that is based on Theorem 2. The strategy aims to reduce the $\tau$ values of reduced graphs as fast as possible. If either $\tau((V \setminus \{v\}, E'))$ or $\tau((V \setminus (V' \cup \{v\}), E''))$ is 0, then by Proposition 1 we can branch on $v$ to a tree, and a minimum vertex cover of a tree may be found in polynomial time. Our strategy can decrease $\tau$ values faster than it can decrease the cardinality of the vertex cover of subgraphs, thus the search tree depth resulting from our approach is less than that of other algorithms. This fact leads to an improved complexity bound.

The next section details the vertex selection algorithm.

## 4   Vertex Selection Procedure

The $VC$ algorithm of Figure 1 recursively reduces a given problem to smaller vertex cover problems by determining or guessing vertices that can be in the minimum cover and reducing the graph and the number $k$ accordingly. At each level of recursion the vertex selection procedure, *select* on line 3 of $VC$, is invoked on input graph $(V, E)$. This procedure has two phases. The first phase might be called a preprocessing step: it finds subgraphs of $(V, E)$ that must contribute a fixed number of vertices to any minimum vertex cover and reduces $(V, E)$ to a smaller graph, either by folding vertices or by eliminating vertices and their incident edges. In the second phase, a branch vertex $v$ is chosen. The procedure

```
        select(V, E) =
1.          v = 0.
2.          Repeat the following until v ≠ 0:
3.              Repeat the following until deg(u) > 2 for all u ∈ V:
4.                  If there is a degree 1 vertex u ∈ V with incident edge {u, w}:
5.                      V := V \ {u}.
6.                      E := E \ {{u, w}}.
7.                  If there is a degree 2 vertex u ∈ V:
8.                      Let edges incident to u be {u, s}, {u, r}.
9.                      For all edges {s, t} ∈ E, t ≠ u, replace s with r.
10.                     V := V \ {s, u}.
11.                     E := E \ {{u, s}, {u, r}}.
12.             If there is a vertex u ∈ V such that deg(u) > 4: set v = u.
13.             If there is a vertex u ∈ V of degree 4:
14.                 cases 1-7 (Figures 3(a-g)): set v = s or as explanation of each case.
15.             If there are only vertices of degree 3:
16.                 cases 1-3 (Figures 3(h-i)): set v as explanation of each case.
17.             Otherwise, V = ∅.
```

**Fig. 2.** Vertex selection procedure

returns $v, V'$, and $E'$ as output. Specifically, the vertex selection procedure is shown in Figure 2.

We describe each operation now except for 1. which needs no explanation. Recall we are interested in analyzing $VC$ for graphs with maximum vertex degree 3. The operations described may result in the replacement of low degree vertices with higher-degree vertices.

From Figure 3.$a$ to $d$ process cases where around $center - vertex$ $u$, there is no triangle and there are degree-3 vertices shared by $N(u)$.

**Case $a$:** In this case, a degree-3 vertex $z$ connects with $u's$ three neighbors. So vertex $z$ has all its neighbors as a subset of vertex $u$'s neighbors.

If $u$ is included in vertex cover, then as stated in [14], at least two of $u$'s neighbors are excluded from vertex cover, so $z$ must be in vertex cover.

If $u$ is not included in vertex cover, then $N(u)$ must be in vertex cover, so $z$ is not in vertex cover too.

From above, we can see that vertex $z$ does not affect minimum vertex cover of $G$, so we can delete vertex $z$ from $G$ and there is still degree-4 vertex left.

**Case $b$:** At least one of $z1, z2$ has degree-3, without losing generality, suppose $deg(z1) = 3$, and $deg(z2)$ can be $\geq 3$.

At first, suppose $deg(s) = 3$, and name its third neighbor $x$.

1. $deg(x) = 3$: If vertex $s$ is included in vertex cover, from $s$ and its three neighbors, 4 extra-degrees can be decreased. And after deleting $s$, $deg(x) = 2$ and can be folded to get big degree vertex or better. After deleting $s$, $deg(z1) = 2$ and can be folded with its two neighbors to be $z1'$. This makes the figure to be isomorphism as Figure 3.h, as will be discussed in **Case** $h$, $z1'$ can be added into vertex cover directly. If $deg(y) = 3$, now $deg(z1') = 4$ and 6 extra-degrees can be decreased. So in total at least 10 extra-degrees are decreased. Or $deg(y) > 3$, now $deg(z1') > 4$ and at least 8 extra-degrees can be decreased, so in total at least 12 extra-degrees can be decreased.

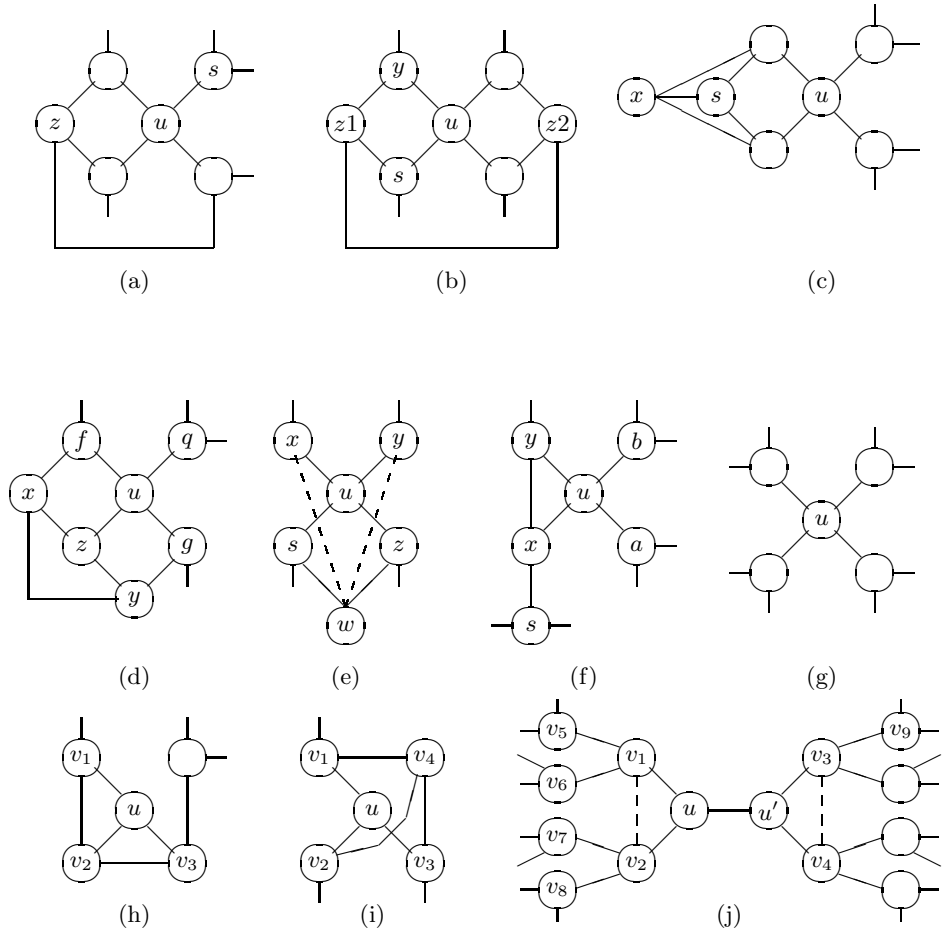

**Fig. 3.** Subgraphs that induce an application of folding and the resulting graph reduction.

If $s$ is not included in vertex cover, $N(s)$ must be included in vertex cover. After deleting $N(s)$, $deg(y) = deg(y) - 2$. So if originally $deg(y) = 3$, now its only neighbor can be added into vertex safely, then at least 14 extra-degrees can be decreased. Or, $deg(y) > 3$ at least 2 extra-degrees can be removed on $y$ and at least 12 extra-degrees can be decreased.

So finally we have branching vector at least as good as as below: 1) $(10, 14)$ with $(G_4, G_3)$ and 2) $(12, 12)$ with $(G_4, G_3)$.

2. $deg(x) = 4$: Do the same thing as subcase $deg(x) = 3$, the only difference is that, when $s$ is included, we can not guarantee degree-4 vertices in sub-instance but two more extra-degrees decreased when $N(s)$ is included in vertex cover. So the branching vector is 1) $(10, 16)$ with $(G_3, G_3)$ and 2) $(12, 14)$ with $(G_3, G_3)$.

Second, suppose $deg(s) = 4$, and its another two neighbors $x, z$. We can do the same thing as $deg(s) = 3$, and it is easy to see that the results are better.

**Case c:** At least one vertex among $x, s$ has degree-3, without losing generality, suppose $deg(x) = 3$. Vertex $s$ can be mapped to vertex $w$ in **Case h**, as will see, $s$ can be added into vertex cover safely. I.e., at least 4 extra-degrees can be decreased without branching.

**Case d:** In this case, there are $deg(x) \geq 3$ and $deg(y) \geq 3$. The analysis is divided into two sub-cases depending on vertex-degree of $z$. Because if $deg(z) = 4$, we can use struction introduced in [14], so next we discuss the condition when $deg(z) = 3$.

If vertex $u$ is included in vertex cover, after deleting $u$, $x, y$ can be added into vertex cover safely. So 8 extra-degrees can be decreased from $u \cup N(u) \cup \{x, y\}$. Here notice that, if originally $deg(f) = 3$ or $deg(g) = 3$, now after deleting $u, x, y$, there is $deg(f) = 1$ or $deg(g) = 1$, so $f/g$'s only neighbor can be added into vertex cover directly, three more extra-degrees can be decreased. If $deg(f/g) > 3$, then more than 1 extra-degree can be decreased on $f/g$. So we have:

1. $deg(f) = deg(g) = 3$ : at least 14 extra-degrees decreased.
2. $deg(f) = deg(4) = 4$ : at least 10 extra-degrees decreased.
3. $deg(f) = 3, deg(g) = 4$ : at least 12 extra-degrees decreased.

If vertex $u$ is excluded from vertex cover, $N(u)$ must be included.

1. $deg(f) = deg(g) = 3$ : at least 12 extra-degrees decreased.
2. $deg(f) = deg(4) = 4$ : at least 16 extra-degrees decreased.
3. $deg(f) = 3, deg(g) = 4$ : at least 14 extra-degrees decreased.

Notice that, if $deg(q) = 3$, then when $u$ is included in vertex cover, after deleting $u$, $deg(q) = 2$. Or $deg(q) > 3$, when $u$ is excluded from vertex cover, two more extra-degrees can be decreased.

So the branching vector is 1) $(14, 12)$ with $(G_4, G_3)$; 2) $(10, 16)$ with $(G_4, G_3)$; 3) $(12, 14)$ with $(G_4, G_3)$; 4) $(14, 14)$ with $(G_3, G_3)$; 5) $(10, 18)$ with $(G_3, G_3)$; 6) $(12, 16)$ with $(G_3, G_3)$.

Until now, we have discussed almost all conditions that around $center - vertex$ $u$, $N[u]$ share degree-3 vertices, but one exception: as displayed in Figure 3.a, vertex $z$ has only its two edges connecting with $N(u)$. In that case, it does not affect sub-instance where $u$ is included in vertex cover. If $N(u)$ is included in vertex cover, after deleting $N(u)$, $deg(z) = 1$ and its only left neighbor can be added into vertex cover directly, so three more extra-degrees can be decreased and the result is much better than no degree-3 vertices shared by $N(u)$.

Next, in **Case** $e$ we discuss conditions that around $center - vertex\ u$, $N[u]$ share degree-4 vertices $w$. Note that, all conditions that: one degree-3 vertex $v$ shared by $N(u)$ and connects with $w$ has been included in **Case b to d**.

**Case** $e$: In this case, $deg(w) = 4$ and Figure 3.$e$ includes several sub-cases: degree-4 vertex $w$ connects with $2/3/4$ $u's$ neighbors.

1. $w$ connects with 2 of $u's$ neighbors $s, z$: Notice that, if one degree-4 vertex as $w$ is not shared by $s, z$, i.e. there are two vertices $w_1, w_2$ connecting with $s, z$ respectively, when $N(u)$ is deleted, from $w_1, w_2$ two extra-degrees can be decreased. But because $deg(w) = 4$, by definition of extra-degree, still 2 extra-degrees can be decreased on $w$ itself. So the results are the same.

   (a) $deg(s) = deg(z) = 4$: Choose to branch on $u$.
   When $u$ is included, 6 extra-degrees can be decreased, and there is degree-4 vertex (e.g. $w$) left in sub-instance.
   If $u$ is not included in vertex cover, $N(u)$ are included in vertex cover. From $N(u)$ and their neighbors, at least 18 extra-degrees can be decreased.
   So the branching vector is $(6, 18)$ with $(G_4, G_3)$.

   (b) $deg(s) = deg(z) = 3$: Choose to branch on $u$. When $u$ is included in vertex cover, after deleting $u$, $deg(s) = 2$ and $deg(z) = 2$ and can be folded with $w$, so a vertex with degree $\geq 6$ can be got. So there is a branching vector $(6, 14)$ with $(G_6, G_3)$.

   (c) $deg(s) = 4, deg(z) = 3$: Name $s$'s another two neighbors $v_1, v_2$.
      i. $deg(v_1) = 3$ or $deg(v_2) = 3$: Choose to branch on $s$.
         If $s$ is included in vertex cover, 6 extra-degrees can be decreased, and after deleting $s$ $deg(v_1) = 2$ or $deg(v_2) = 2$ which can be folded.
         When $s$ is not included in vertex cover, so $N(s)$ are in vertex cover and at least 18 extra-degrees can be decreased.
         So the branching vector is at least $(6, 18)$ with $(G_4, G_3)$.
      ii. $deg(v_1) = deg(v_2) = 4$: Choose to branch on $u$.
         If $u$ is included in vertex cover, 6 extra-degrees can be decreased.
         When $s$ is not included in vertex cover, so $N(s)$ are in vertex cover. Notice that after deleting $N(s)$, $deg(s) = 1$ so its another neighbor can be added into vertex cover, three more extra-degrees can be decreased. In a total 24 extra-degrees can be decreased.
         So the branching vector is at least $(6, 24)$ with $(G_3, G_3)$.

2. $w$ connects with 3 of $u's$ neighbors $s, z, x$ Choose to branch on $u$. When $u$ is excluded from vertex cover, after deleting $N(u)$, $deg(w) = 1$, and its only neighbor can be added into vertex cover, so from $w$, at least 5 extra-degrees can be decreased. It is easy to see that the result is better than $w$ connects with 1 or 2 neighbors of $u$. And we can analyze this depending on $deg(s), deg(z), deg(x)$ as last subcase.

3. $w$ connects with 4 of $u's$ neighbors Similar as $Case\ a$, we can delete $w$ from graph $G$.

In **Case** $f$, we discuss conditions where around $center - vertex$ $u$, there are triangles. Because of struction [14] and $deg(u) = 4$, we can assume in triangle there is no degree-3 vertex.

**Case** $f$:

In $u's$ neighbors, every vertex like $x$ has at least one neighbor $s \notin \{u \cup N(u)\}$, i.e. no vertex in $N(u)$ can be in more than two triangles, or it is dominated by $u$ and $u$ can be added into vertex cover directly.

In this case, as displayed in figure, we have $deg(x) = deg(y) = 4$. So it is easy to see that if we branch on $u$, more extra-degrees can be decreased than **Case** $g$, and the only thing we need to consider is *whether there are degree-4 vertices in sub-stances*. Then in Figure 3.$f$, if $deg(a) = 3$ or $deg(b) = 3$, it can be analyzed the same as in Figure 3.$g$. Or all vertices in $N(u)$ have degree $\geq 4$, because one vertex in $N(u)$, say $y$, can not connect with all other vertices in $N(u)$. For example, $y$ and $a$ are not connected by one single edge, then we can choose to branch on vertex $a$, i.e. choose $a$ as the $center - vertex$. Easy to see when $a$ is included in vertex cover, there are vertices, e.g. $y$, in the reduced graph. In this subcase, the conditions when branched vertex is excluded from vertex cover are the same as Figure 3.$g$.

**Case** $g$: When a degree-4 vertex $u$ is included in vertex cover, if the reduced subgraph has no degree-4 vertex, the only possibilities are that i) vertices in $N(u)$ have degree $\geq 4$; or ii) after deleting $u$, the generated degree-2 vertices are connected pair by pair and so their number must be even. For first condition, as described in **Case** $f$, we can choose one vertex in $N(u)$ to branch on. For the later condition, that means there are triangles around $u$ and at least one triangle contains degree-4 and degree-3 vertices, so can be applied with struction rule. The discussion here means that we only need to think the condition when branched vertex $u$ excluded.

If $u$ is included in vertex cover, from $u \cup N(u)$, 6 extra-degrees can be decreased, and more than one *nonadjacent* degree-2 vertex can be got; If $u$ is excluded from vertex cover, $N(u)$ must be in vertex cover, then from $N(u)$ and $N(u)'s$ neighbors, at least 14 extra-degrees can be decreased. If we assume that when $u$ is excluded from vertex cover, there are degree-4 vertices in corresponding subgraph, we have a branch number as $(6, 14)$ with $(G_4, G_4)$.

So now we consider conditions when $u$ excluded from vertex cover, there is no degree-4 vertex in corresponding subgraph. That are several possibilities list as below and one possibility displayed in Figure 4.$a$:

1. Vertices $w_1, ..., w_8$ have degree $\geq 4$.
2. Vertices $w_1, ..., w_8$ have degree 3 but they are connected pair by pair just as in Figure 4.$a$.
3. A mix conditions of the above two.

If there are many degree-4 conditions, it is easier to decrease more extra-degrees, that is because we can choose the most beneficial one to branch on. For example, if in $w_1, ..., w_8$ there are degree-4 vertices, we can choose to branch on them instead of $u$. If all vertices $w_1, ..., w_8$ have degree $\geq 4$, we can find a vertex $w_i$ when it is excluded from vertex cover, at leat one vertex in $w_1, ..., w_8$ is not affected and will be the degree-4 vertex in subgraph. If no such vertex $w_i$ exists, that means around one vertex in $w_1, ..., w_8$, there are degree-4 vertices and when it is excluded from vertex cover, more extra-degrees can be decreased. If only part of $w_1, ..., w_8$ have degree-4, we can see that the actions applied on Figure 4.$a$ can be still used.

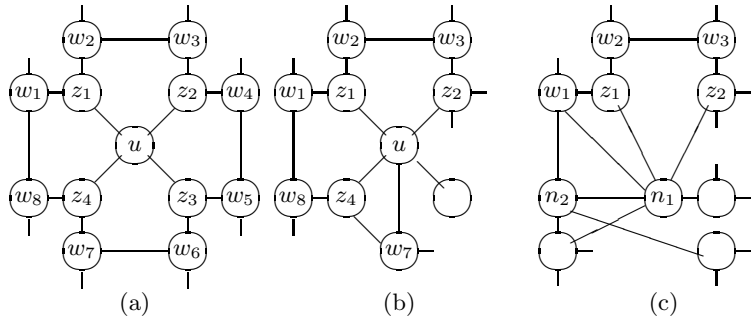

**Fig. 4.** Subcases of Figure 3.$g$

In Figure 4.$a$, choose to branch on $w_5$. Here not only we will introduce how to process this case and also introduce how to use struction.

If $w_5$ is included in vertex cover, from its neighbors 4 extra-degrees can be decreased. After deleting $w_5$, $deg(z_3) = deg(w_4) = 2$ and can be folded to be degree-4 vertices and the subgraph is displayed in Figure 4.$b$. Now we can apply struction on vertex $z_4$ and its neighbors to get Figure 4.$c$, and it's easy to see that 2 extra-degrees are increased. In Figure 4.$b$, vertices $z_1, w_1, n_1$ in the triangle and $deg(z_1) = 3$ so struction can be applied again. After struction, 2 extra-degrees are increased and one degree-7 vertex generated whose neighbors contain one degree-5 vertex and four degree-4 vertices. By branching on the degree-7 vertex we can have a branching vector $(12, 34)$ with $(G_4, G_3)$. Summing up, this branching has a branching vector $(12+4-2-2, 34+4-2-2) = (12, 34)$ with $(G_4, G_3)$.

If $w_5$ is excluded from vertex cover, then $N(w_5)$ must be in vertex cover, and so 10 degrees can be decreased and in the subgraph $z_2$ can be folded to generate one degree-4 vertex.

So finally, out branching vector is $(34, 12, 10)$ with $(G_3, G_4, G_4)$.

**Case** $h$: Vertex $v_2$ is dominated by $u$, so $u$ can be added into vertex cover directly.

**Case** $i$: By $general - folding$, this sub-figure can be folded.

Also, if $u$ is included in vertex cover, at least two vertices in $v_1, v_2, v_3$ are not in vertex cover, so $v_4$ is in vertex cover; if $u$ is not included in vertex cover, vertices $v_1, v_2, v_3$ are in vertex cover, so $v_4$ is not in vertex cover neither. So we can delete $u$ or $v_4$ from graph $G$.

In the last case we need to prove that in worst case, we have a branching vector either $(4, 10)$ to generate subgraphs both containing degree-4 vertices, I.e.

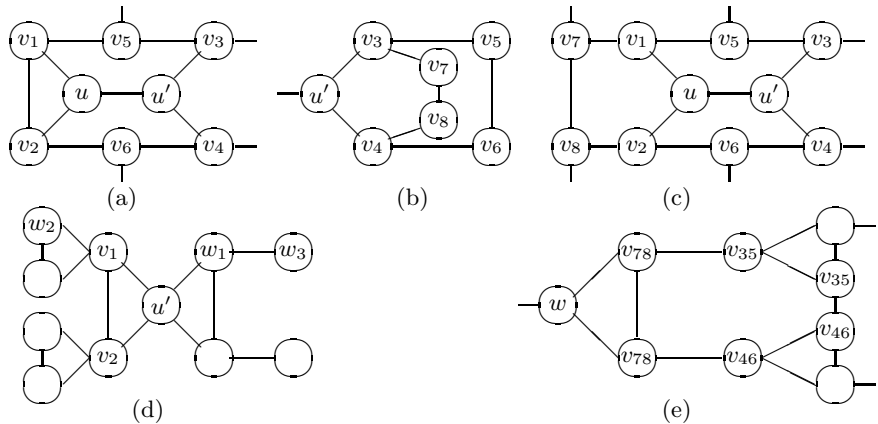

**Fig. 5.** Subcases of Figure 3.*j*

the worst branching vector will be $(4, 10)$ with $(G_4, G_4)$; or at least as good as $(6, 14)$ with $G_3, G_3$).

**Case *j*:** There are several subcases in Figure 3.*j* where the graph $G$ is 3-regular.

1. There are triangles in $G$: If around vertex $u$, there are more than one triangle, then it is **Case *h***. So there can be only one triangle around $u$ as displayed in Figure 3.*j*. In Figure 3.*j*, choose to branch on vertex $u'$. Here we can assume there is no edge between $v_3$ and $v_4$, or we can choose one vertex in $N(v_3) \cup N(v_4) \setminus \{u'\}$, and it is easy to see that the result is better than branching on vertex $u'$. From **Case *i***, we only need to think conditions as below:
   (a) Except $u'$, two vertices in $u, v_3, v_4$ have one common neighbor.
   (b) Vertices in $u, v_3, v_4$ have no common neighbor except $u'$.
   At first, we assume the later condition and then the prior ones.
   If $u'$ is included in vertex cover, 4 extra-degrees can be decreased and after deleting $u'$, $deg(u) = 2$, so $v_1, v_2$ can be added into vertex cover. In a total, 8 extra-degrees can be decreased. If $u'$ is excluded from vertex cover, then $N(u')$ must be in vertex cover and hence 10 extra-degrees decreased.
   When $u'$ is included in vertex cover, the only possibilities that the subgraph has no degree-4 vertices are: i) $v_1, u'$ (or $v_2, u'$) have common neighbors; ii) neighbors of $u', v_1, v_2$ are connected by edges. As discussed before, in condition *i*, after deleting $u', v_1, v_2$, their common neighbors have degree 1 and so they can be folded to add their left neighbors into vertex cover, hence more extra-degrees can be decreased. Because of this, we only consider condition *ii* displayed as Figure 5.*a*. Then we choose to branch on vertex $v_4$. If $v_4$ is included in vertex cover, 4 extra-degrees can be decreased and surely after deleting $v_4$ $u'$ can be folded to get degree-4 vertex. If $v_4$ is not included in vertex cover, then $N(v_4)$ must be in vertex cover, and also $v_1, v_2$ can be added into vertex cover. In a total, at least 12 extra-degrees can be decreased, and we get 5 degree-2 vertices and so can guarantee at least one degree $\geq 4$ vertices generated. Hence the branch vector is $(4, 12)$ with $(G_4, G_4)$ and is better than $(4, 10)$ with $(G_4, G_4)$.

When $u'$ is excluded from vertex cover, if the subgraph has no degree-4 vertex, then neighbors of $v_3, v_4$ must connected pair by pair.

(a) $v_1$ or $v_2$ is connected with neighbors of $v_3, v_4$ except $u'$ by a single edge. For example, there is one edge between $v_1$ and $v_9$. Then we can choose to branch on $v_9$'s third neighbors except $v_1, v_3$. Then when the branched vertex is included vertex cover, after deleting it, $deg(v_9) = 2$ and can be folded. Then $v_1$ can be added into vertex cover. Finally, the branch vector can be at least $(8, 14)$ with $(G_3, G_3)$.

(b) Neighbors of $v_3, v_4$ connected pair by pair as displayed in Figure 5.b. Choose to branch on vertex $v_3$. When it is included in vertex cover, 4 extra-degrees can be decreased, and can generate three *nonadjacent* degree-2 vertices which can be folded to get degree $\geq 4$ vertices. If $v_3$ is excluded from vertex cover, $N(v_3)$ must be in vertex cover. After deleting them, $deg(u) = 2$ so $v_1, v_2$ can be added into vertex cover. 14 extra-degrees can be decreased and 7 degree-2 vertices produced. So at least one degree $\geq 4$ vertices in subgraph. The branch vector is at least $(4, 14)$ with $(G_4, G_4)$.

2. No triangle in $G$: There is no edge between $v_1, v_2$ or $v_3, v_4$. Choose to branch on $u$. If $w$ is included in vertex cover, 4 extra-degrees can be decreased and get three *nonadjacent* degree-2 vertices, so they can be folded to get degree-4 vertices.

If $N(u)$ are included in vertex cover, at least 10 extra-degrees can be decreased. The only way after deleting $N(u)$ no degree $\geq 4$ vertices generated is that $v_3, v_4, ..., v_8$ are connected with each other pair by pair or $u', v_1, v_2$ share common neighbors. Here notice that we can choose arbitrary vertex to branch to avoid that the subgraphs have no degree-4 vertex. For example, if one vertex $v$ is shared as common neighbor of $v_1, v_2$, then we can choose to branch on $v$'s third neighbor. Then when that vertex is excluded from vertex cover, its neighbors including $v$ must be in vertex cover, then after deleting them $v$'s another two neighbors $v_1, v_2$ have degree-2. Because there is no triangle, there is no edge between $v_1, v_2$. If $v_1, v_2$ share third vertex as common neighbor except $v, u$, then they are dominated by each other as in Figure 3.i. So we can guarantee there are degree $\geq 4$ vertices in subgraphs. So the only possibility is displayed as in Figure 5.c, where includes *pentagon* as a part. Remind that we can choose arbitrary vertex to branch, so this means every vertex is surrounded by those *pentagons*. Choose to branch on vertex $u$.

If $u$ is included in vertex cover, 4 extra-degree decreased and after deleting it there are three *nonadjacent* degree-2 vertices including $u', v_5$. Folding them we get three connected triangles including two degree-3 vertices and one degree-4 vertex as drew in Figure 5.d. Here we name the new vertices from the names $u', v_1, v_2$ which they are folded from respectively. $w_1, w_2$ are neighbors of $v_3, v_5$ respectively. $w_1, w_2$ must share one common neighbor named $w_3$. Because $v_1, u$ as neighbors of $v_5, u'$ are connected by one single edge, then $w_2$ must connect with one $w_1$'s neighbor except $v_3$ or we can choose to branch on $v_3$, and when $v_3$ is excluded from vertex cover, there are degree $\geq 4$ vertices in subgraph which leads a contradiction with our assumption. So in Figure 5.*d*, $w_3$ connects with $w_1$ and $w_2$, if choose to branch on it, we can decrease at least $(16, 10)$ extra-degrees.

If $u$ is not included in vertex cover, 10 extra-degrees decreased and after deleting $N(u)$, we get 6 degree-2 vertices which connect pair by pair. Similar as analysis of existence of $w_3$, by folding them, no degree $\geq 4$ vertices generated but gets part of the subgraph showed in Figure 5.*e*. For ease to track the folding, we use $v_{35}$ to represent the two vertices coming from the connected pair of vertices $v_3, v_5$ and their two neighbors, and the same for $v_{46}, v_{78}$. Choose to branch on vertex $w$'s third neighbor, we can decrease more than $(14, 10)$ extra-degrees.

Finally, we get a branch vector $(24, 20, 20, 14)$.

## 5 Complexity Analysis

Next we use the same method as in [4] to calculate branching number which indicate the size of branched tree of subgraphs. Use $F_i(x)$ to represent the branching number when branching on a graph with degree-$i$ vertices and $x$ denote the $\tau$ value of the initiate graph $G$. By Corollary 1, decreasing 2 extra-degrees can decrease $\tau$ by 1.

By exhaust all combinations of branching vector in Section 4, we can calculate out the worst branching number $1.15855$ easily. Here we give some examples to show how to count for it.

In Section 4 **Case** *b*, there is branching vector $(10, 14)$ with $(G_4, G_3)$, so we have a branching vector $(5, 7)$ with $(G_4, G_3)$ for $\tau$ value and from now, all branching vector is discussed on $\tau$. From Section 4 we know that by branching on a graph with only degree-3 vertices, the worst branching vector is $(2, 5)$ with $(G_4, G_4)$ for $\tau$. So we have $(5, 7+2, 7+5) = (5, 9, 12)$ with $(G_4, G_4, G_4)$ whose branching number is $1.1451$ which is better than $1.15855$.

In Section 4 **Case** *g*, there is branching vector $(6, 14)$ with $(G_4, G_4)$ whose branching number is $1.15855$ which is also the worst branching number. Also there is branching vector $(17, 6, 5)$ with $(G_3, G_4, G_4)$. Again, we can have a new branching vector $(22, 19, 6, 5)$ with $(G_4, g_4, G_4, G_4)$ whose branching number is $1.1574$ which is better than $1.15855$. Thus we have a proposition as below:

**Proposition 2.** *On a graph $G = (V, E)$ containing degree $\geq 4$ vertices, can branch on vertices to get a group of subgraphs which all contain $\geq 4$ vertices and the branch number of decreasing $\tau$ is not worse than $(3,7)$'s value $1.15855$.*

**Theorem 3.** *Algorithm in Figure 1 can find a k vertex cover with complexity bounded by $O(1.15855^k)$.*

*Proof.* From [10,5], a vc-3 problem $(G, k)$ can be transformed in running time $O(k^{3/2})$ to a new vc-3 problem $(G_1, k_1)$. And $G_1$ has the property that $n \leq 2 \cdot k_1$,

$k_1 \leq k$. Also $G_1$ has a vertex cover with cardinality of $k_1$ if and only if $G$ has a vertex cover with cardinality of $k$. Since $n \leq 2 \cdot k_1$ and by Corollary 1, then there are at most $floor(\frac{2k}{2})+1 = k+1$ real cycles. On a given graph $G$, after constant times of branching, all the subgraphs will contain degree $\geq 4$ vertices. From Proposition 2, the worst branching number is 1.15855. Then the complexity to remove all real-cycles is $O(1.15855^k)$. Because a graph without cycle is a tree graph and the minimum vertex cover problem can be solved in polynomial time, the complexity bound of the algorithm is $O(1.15855^k)$.

**Corollary 2.** *Algorithm in Figure 1 can be improved to find $k$ vertex cover with complexity bound $O(1.1504^k)$.*

*Proof.* By using method introduced in [5,13], we can have a new complexity as $O(1.15855^{k-\alpha k} + 16^{\alpha k}1.15855^{\alpha k})$. Choose $\alpha = 0.04799$ so that $1.15855^{k-\alpha k} = 16^{\alpha k}1.18156^{\alpha k}k$, then the new complexity is $O(1.15855^{k-0.04799k}) = O(1.1504^k)$.

## 6  Conclusion and Next Step of Work

In this paper, we give a new algorithm for finding the minimum vertex cover problem on graph with maximum degree 3 where $k$ is the cardinality of vertex cover ($vc-3$ problem). In our algorithm, we try to choose vertex to branch to reduce the graph to be cycle-free graph, because it is known that if the graph is a tree like graph, its minimum vertex cover can be found in polynomial time. Also, we use a new way on analysis to keep on branch on subgraphs contain degree $\geq 4$ vertices. With there new techniques, we get a new complexity bound $O(1.1504)$ which is better than current complexity bound. We notice that after branching of some vertices, there may be many degree $\geq 4$ vertices and a better branching number of decreasing extra-degrees can be achieved leading to a lower complexity bound. This will be our next step of work in the future.